\documentclass[twoside,11pt]{article}

\usepackage{jmlr2e}
\usepackage{shortcuts}

\ShortHeadings{\emph{ruptures}: change point detection in Python}{Truong, Oudre and Vayatis}
\firstpageno{1}

\begin{document}

\title{\emph{ruptures}: change point detection in Python}

\author{\name Charles Truong \email truong@cmla.ens-cachan.fr \\
	\name Nicolas Vayatis \email vayatis@cmla.ens-cachan.fr \\
	\addr CMLA, ENS Cachan,	CNRS, Université Paris-Saclay \\
	94235, Cachan, France\\
	COGNAC G, University Paris Descartes, CNRS \\
	75006 Paris, France
	\AND
	\name Laurent Oudre \email laurent.oudre@univ-paris13.fr\\
	\addr L2TI, University Paris 13 \\
93430 Villetaneuse, France}%

\editor{}

\maketitle

\begin{abstract}
	\ruptures{} is a Python library for offline change point detection.
	This package provides methods for the analysis and segmentation of non-stationary signals.
	Implemented algorithms include exact and approximate detection for various parametric and non-parametric models.
	\ruptures{} focuses on ease of use by providing a well-documented and consistent interface.
	In addition, thanks to its modular structure, different algorithms and models can be connected and extended within this package.
\end{abstract}


\section{Introduction}
Change point detection is the task of finding changes in the underlying model of a signal.
This subject has generated important activity in statistics and signal processing~\citep{Lavielle2005, Jandhyala2013, Haynes2017}.
Modern applications in bioinformatics, finance, monitoring of complex systems have also motivated recent developments from the machine learning community~\citep{Vert2010,Lajugie2014,Hocking2015}.\\
We present \ruptures{}, a Python scientific library for multiple change point detection in multivariate signals. 
It is meant to answer the growing need for fast exploration,  by non-specialists, of non-stationary signals.
In addition, we expect that removing the cost of reimplementation will facilitate composition of new algorithms.
To that end, \ruptures{} insists on an easy-to-use and consistent interface.
Implementation is also modular to allow users to seamlessly plug their own code.\\
To the best of the authors' knowledge, \ruptures{} is the first Python package dedicated to multiple change point detection.
Most related softwares are implemented in R~\citep{Erdman2007,James2014,Killick2014,Ross2015,Fryzlewicz2017,Haynes2017,Chakar2017}.
However, few provide more than one algorithm, and even fewer can be applied to detect changes other than mean shifts.
On the other hand, \ruptures{} contains several standard methods as well as recent contributions, most of which are not available elsewhere (in Python or R).
Our work encompasses most packages and provides a unique framework to run and evaluate all algorithms.\\
In the following, we quickly describe the change point detection framework.
Then the main features of the library are detailed.

\section{Change point detection framework}
\label{sec:cpd}
In the offline (or retrospective) change point detection framework, we consider a non-stationary random process $y = \{y_1,\dots,y_T\}$ that takes value in $\RR^d$ ($d\geq1$).
The signal $y$ is assumed to be piecewise stationary, meaning that some characteristics of the process change abruptly at some unknown instants $\sta{t_1}<\sta{t_2}<\dots<\sta{t_K}$.
Change point detection consists in estimating those instants when a particular realization of $y$ is observed.
Note that the number of changes $K$ is not necessarily known.\\
Most estimation methods adhere to or are an approximation of a general format where a suitable contrast function $V(\cdot)$ is minimized~\citep{Jandhyala2013,Lavielle2005}.
Usually, it is written as a sum of segment costs:
\begin{equation}
	V(\ttt, y) := c(\{y_t\}_{1}^{t_{1}}) + c(\{y_t\}_{t_1+1}^{t_{2}}) +\dots+ c(\{y_t\}_{t_i+1}^{t_{i+1}}) +\dots
\end{equation}
where $\ttt=\{t_1,t_2,\dots\} $ denotes a set of change point indexes and $c(\cdot)$ denotes a cost function that takes a process as input and measures its goodness-of-fit to a specified model.
The contrast $V(\cdot)$ is the total cost associated with choosing a particular segmentation $\ttt$.
Change point detection amounts to solving the following discrete optimization problem:
\begin{equation}
	\min_{\ttt} V(\ttt, y) + \text{pen}(\ttt)
	\label{eq:mincost}
\end{equation}
where $\text{pen}(\ttt)$ is a regularizer on the value of the partition $\ttt$.
Methods from the literature essentially differ by 1) the constraints they add to this optimization problem (fixed dimension of $\ttt$, penalty term, cost budget, etc.), 2) how they search for the solution (exact or approximate resolution, local or sequential, etc.) and 3) the cost function $c(\cdot)$ they use (which is related to the type of change).

\section{Library overview}
\begin{figure}[t]
	\centering
	\includegraphics[width=0.8\textwidth]{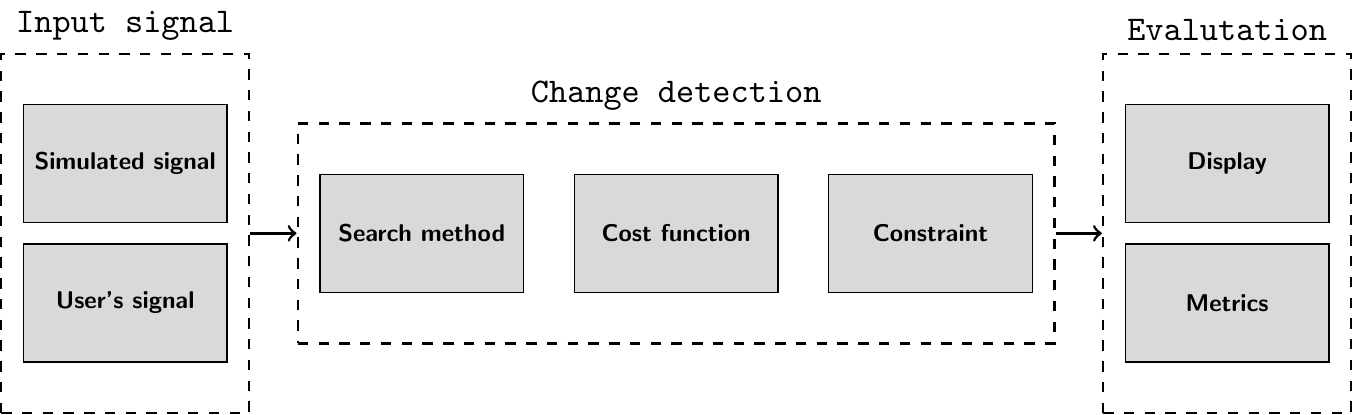}
	\caption{Schematic view of the \ruptures{} package.}
	\label{fig:diagram}
\end{figure}

A basic flowchart is displayed on Figure~\ref{fig:diagram}.
Each block of this diagram is described in the following brief overview of \ruptures{}' features.
More information can be found in the related documentation (see link to source in Section~\ref{sec:info}).
\subsection{Main features}
\begin{itemize}
	\item\textbf{Search methods}\enspace
	Our package includes the main algorithms from the literature, namely dynamic programming, detection with a $l_{0}$ constraint, binary segmentation, bottom-up segmentation and window-based segmentation. 
	This choice is the result of a trade-off between exhaustiveness and adaptiveness.
	Rather than providing as many methods as possible, only algorithms which have been used in several different settings are included.
	In particular, numerous ``mean-shift only'' detection procedures were not considered.
	Implemented algorithms have sensible default parameters that can be changed easily through the functions' interface.
	\vspace*{-0.5em}
	\item\textbf{Cost functions}\enspace 
	Cost functions are related to the type of change to detect. 
	Within \ruptures{}, one has access to parametric cost functions that can detect shifts in standard statistical quantities (mean, scale, linear relationship between dimensions, autoregressive coefficients, etc.) and non-parametric cost functions (kernel-based or Mahalanobis-type metric) that can, for instance, detect distribution changes~\citep{harchaoui2007retrospective,Lajugie2014}.
	\vspace*{-0.5em}
	\item\textbf{Constraints}\enspace
	All methods can be used whether the number of change points is known or not.
	In particular, \ruptures{} implements change point detection under a cost budget and with a linear penalty term~\citep{Killiack2012a, Maidstone2017}.
	\vspace*{-0.5em}
	\item\textbf{Evaluation}\enspace 
	Evaluation metrics are available to quantitatively compare segmentations, as well as a display module to visually inspect algorithms' performances.
	\vspace*{-0.5em}
	\item\textbf{Input}\enspace 
	Change point detection can be performed on any univariate or multivariate signal that fits into a \emph{Numpy} array.
	A few standard non-stationary signal generators are included.
	\vspace*{-0.5em}
	\item\textbf{Consistent interface and modularity}\enspace Discrete optimization methods and cost functions are the two main ingredients of change point detection.
	Practically, each is related to a specific object in the code, making the code highly modular: available optimization methods and cost functions can be connected and composed.
	An appreciable by-product of this approach is that a new contribution, provided its interface follows a few guidelines, can be integrated seamlessly into \ruptures{}.
	\vspace*{-0.5em}
	\item\textbf{Scalability}\enspace
	Data exploration often requires to run several times the same methods with different sets of parameters.
	To that end, a cache is implemented to keep intermediate results in memory, so that the computational cost of running the same algorithm several times on the same signal is greatly reduced.
	We also add the possibility for a user with speed constraints to sub-sample their signals and set a minimum distance between change points. 
\end{itemize}
\subsection{Availability and requirements}
\label{sec:info}
The \ruptures{} library is written in pure Python and available on Mac OS X, Linux and Windows platforms.
Source code is available from \liencode{} under the BSD license.
We also provide a complete documentation that includes installation instructions, explanations with code snippets on advance use (\liendoc{}).\\
Implementation relies on \emph{Numpy} as the base data structure for signals and parameters and \emph{Scipy} for efficient linear algebra and array operations.
The \emph{Matplotlib} library is recommended for visualization.
Unit tests (through the \emph{Pytest} library) are provided to facilitate the validation of new pieces of code.
\subsection{Illustrative example}
\begin{figure}[t]
	\centering
	\subfloat[Python code.]{
		\includegraphics[width=0.4\textwidth]{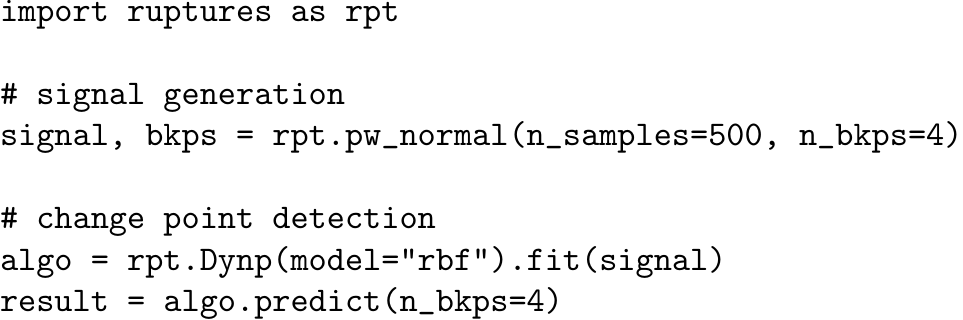}
		\label{fig:code}
	}
	\subfloat[Top and middle: simulated 2D signal; regimes are highlighted in alternating gray area. Below: scatter plots for each regime type.]{
		\includegraphics[width=0.4\textwidth]{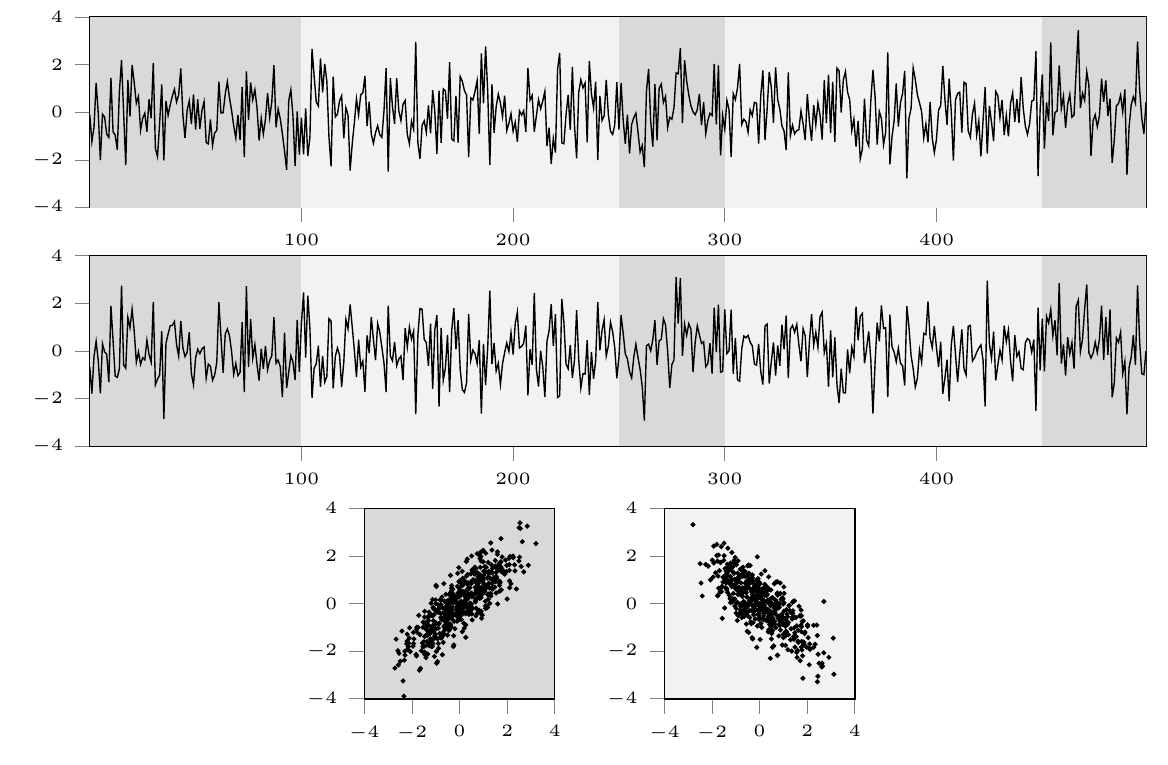}
		\label{fig:signal}
	}
	\caption{Illustrative example.}
	\label{fig:example}
\end{figure}%
As an illustrative example, we perform a kernel change point detection on a simulated piecewise stationary process~\citep{harchaoui2007retrospective}.
In a nutshell, this method maps the input signal onto a high-dimensional Hilbert space $\mathcal{H}$ through a kernel function (here, we use the radial basis function) and searches for mean shifts.\\
First, random change point indexes are drawn and a 2D signal of \iid\ centred normal variables with changing covariance matrix is simulated (Figure~\ref{fig:signal}).
The algorithm's internal parameters are then fitted on the data.
The discrete minimization of the contrast function is performed with dynamic programming and the associated estimates are returned.
The related code lines are reported on Figure~\ref{fig:code}.\\
It is worth mentioning that only a few instructions are needed to perform the segmentation.
In addition, thanks to \ruptures{}, variations of the kernel change point detection can be easily carried out by changing a few parameters in this code.
\section{Conclusion}
\ruptures{} is the most comprehensive change point detection library.
Its consistent interface and modularity allow painless comparison between methods and easy integration of new contributions.
In addition, a thorough documentation is available for novice users.
Thanks to the rich Python ecosystem, \ruptures{} can be used in coordination with numerous other scientific libraries
\vspace*{-2em}
\acks{This work was supported by a public grant as part of the Investissement d'avenir project, reference ANR-11-LABX-0056-LMH, LabEx LMH.}
\bibliography{biblio}

\end{document}